# Interaction of superconductor with magnetic sheath as a way for improvement of critical current in $MgB_2$/Fe superconductor

J. Horvat

Institute for Superconducting and Electronic Materials, University of Wollongong, NSW 2522, Australia

**Abstract**

Magnesium diboride superconducting wires give the largest critical current density ($J_c$) when produced with iron sheath. Because iron is ferromagnetic, it is expected to improve the field dependence of $J_c$ by shielding of the external field for low magnetic fields. However, transport and magnetic measurements of $J_c$ reveal that $J_c$ in $MgB_2$/Fe is improved far beyond the effect of simple magnetic shielding. The transport measurements in external field show that $J_c$ initially decreases with the field. This is followed by an increase for intermediate fields and again a decrease for high fields, resembling the "peak effect". The value of $J_c$ in the field range of this peak effect is higher than the $J_c$ without iron sheath. The field range of improved $J_c$ widens with decreasing the temperature, shifting to the higher values of the field. The explanation of this phenomenon is suggested in terms of a model predicting the occurrence of overcritical state, as a result of interaction between partly vortex filled superconductor and a magnet. In this model, the currents are pushed into vortex-free volume of the superconductor, effectively increasing its value of loss-free current. The occurrence of the overcritical state is supported by magnetic measurements of $J_c$.

## 1. Introduction

High-temperature superconductors (HTS) offer a chance of a widespread introduction of superconductivity in practical applications. This is because of high value of their critical temperature, enabling the use of inexpensive liquid nitrogen to maintain the superconductivity. However, HTS have two major drawbacks that severely limit their critical current density ($J_c$): *grain connectivity* and *vortex pinning* [1,2]. The problem of grain connectivity was solved quite successfully with Bi2223 superconductor. Unfortunately, vortex pinning is rather weak for Bi2223 because of its large superconducting anisotropy. This results in a rather strong decrease of $J_c$ with magnetic field, making Bi2223 unsuitable for high field applications. Less anisotropic HTS, like Y123, have much stronger vortex pinning and can be used in large magnetic fields. However, the grain connectivity of these HTS is poor and they still cannot provide high $J_c$ on large scale. While Bi2223/Ag tapes are being gradually introduced into low-field applications, such as transformers and electrical power transport, there is still no practical HTS capable of producing high $J_c$ in large magnetic fields.

Discovery of superconductivity in $MgB_2$ offered new alternatives[3]. The critical temperature of $MgB_2$ is 39 K and liquid nitrogen is not suitable for maintaining its superconductivity. However, modern cryocoolers can readily maintain 20K, which would be well suited for applications of $MgB_2$. Unlike HTS, this superconductor has no problem with grain connectivity[4,5] and costs of its production are many times lower. The field dependence of $J_c$ for pure $MgB_2$ is much better than for Bi2223, but it is still not good enough for practical applications. One of the reasons for this is its low value of the upper critical field, about 18T [4]. Because of this, much effort is under way to improve its vortex pinning[6,7,8].

Because $MgB_2$ is brittle, it made sense to make the superconducting wires with this material in the same way as Bi2223, by sheathing $MgB_2$ with a ductile metal. Various metals and alloys were tried, including silver [9,10], copper [11, 9], stainless steel[12], nickel[13] and iron[14], for example. The iron-sheathed wires gave the best values of $J_c$ [6, 14]. This compatibility of $MgB_2$ with iron opened up a possibility to use magnetic shielding for improving the AC loss and field dependence of $J_c$ of $MgB_2$/Fe wires.

Magnetic shielding is expected to provide a strong improvement of AC loss in multifilamentary $MgB_2$/Fe wires. This concept has been proposed for multifilamentary HTS by Majoros et al. [15]. They showed that the transport AC loss in a multifilamentary superconducting tape, in which the filaments are surrounded by a medium of high magnetic permeability, depends on the shape of the filaments. Their model takes into account only re-distribution of magnetic field due to the magnetic shielding. It was assumed in the model that, the currents in the superconductor are not affected by the magnetic surroundings. Experiments on a Bi2223/Ag superconducting tape, coated with iron powder on the surface of the tape, showed a decrease of the critical current by 20% [15]. However, this coating partly screened out the field produced by a neighbouring tape carrying the current. These experiments proved the viability of the idea of magnetic shielding for lowering the AC loss. However, this method is not suitable for practical applications because of associated discontinuities in the layer of iron powder, resulting in poor shielding effect. Iron cannot be used as a sheath material for Bi2223 tapes because of their chemical incompatibility. However, iron is the material of choice for production of $MgB_2$/Fe wires.

The magnetic shielding itself is not expected to provide a breakthrough in improving the field dependence of $J_c$. The shielding effectiveness is limited by magnetic susceptibility of iron and thickness of the iron sheath. The latter cannot be made too thick for practical reasons and it is not viable to shield the superconductor from the fields higher than about 1 T. Surprisingly, transport measurements of $J_c$ for $MgB_2$/Fe wires in magnetic field showed that the improvement of $J_c$ goes way beyond the effect of mere magnetic shielding[16, 17]. The observed effect is also expected to influence the suppression of AC loss by the iron sheath.

This contribution describes the effect leading to the unusually strong improvement of field dependence of transport $J_c$ of $MgB_2$ by the iron sheath. A possible explanation is given in terms of an interaction between the superconductor and magnet.

**2. $MgB_2$/Fe wires**

Round $MgB_2$/Fe wires were prepared by filling an iron tube with fine magnesium and amorphous powders[18]. Each of the powders was 99% pure. They were mixed with the stoichiometric ratio of magnesium to boron of 1:2, respectively. The diameter of the iron tube was 10mm. The filled tube was drawn into a wire of outer diameter of typically 1.5 mm. The inner diameter, containing the mixed powders, was typically 0.8 mm.

The wire was cut into 2 cm long samples, which were sealed into a short iron tube. This tube was then heated to 840°C in flowing high-purity argon. The temperature of the samples in the tube reached its maximum value within 3 minutes. The samples remained at this temperature for 10 minutes, during which time $MgB_2$ was fully formed in the wire. They were then quenched to room temperature in air.

All the samples of $MgB_2$/Fe wires consisted of more than 90% of $MgB_2$, as obtained from x-ray diffraction patterns. The other phases were MgO and traces of unreacted boron and magnesium. The core of the wires consisted of superconducting grains smaller than 1 μm. The density of the core was 1.6 g/cm$^3$, only about 65% of the density of the $MgB_2$ crystal. Even so, the grains were well connected, providing high values of $J_c$. The critical temperature ($T_c$) was obtained form the measurements of AC susceptibility. The onset of the screening in the real part of ac susceptibility occurred for all samples at 38.2 K, with variation between the samples smaller than 0.5K.

## 3. Measurement of $J_c$

The values of $J_c$ for these wires are of the order of $10^5$ A/cm$^2$ above 20K, which requires the use of the currents of several hundreds of amperes to perform the measurements of $J_c$. These currents would produce unacceptably high heating at the contacts of the connecting current-leads to the samples. Because of this, a very short pulse of the current was applied to the sample for each measurement. The duration of the pulse was several milliseconds. The total heat produced in this pulse was small enough not to raise the sample temperature before the measurements were taken. The value of $J_c$ was obtained by dividing the critical current ($I_c$) by the cross-sectional area of the superconducting core. $I_c$ was obtained from voltage-current (VI) characteristics, as the value of the current at which a sudden increase of the voltage was observed.

VI characteristics were measured by a four-probe method. The current leads were soldered at the sample ends. Voltage leads were soldered in the middle of the sample, at a distance of 1 cm from each other. The current was produced by discharging a capacitor through the sample and a non-inductive resistor of resistance 0.01 Ω, connected in series to the sample. A solenoid made of very low-resistance wire was also connected in series to the sample, capacitor and resistor. The role of the solenoid was to prolong the initial increase of the current, during which the measurement was taken. Its inductance was adjusted so that the maximum current was achieved in about a millisecond after the beginning of the pulse of the current. The voltage drop across the resistor was used to obtain the value of the instantaneous current through the sample. The voltage signal was first amplified 100 times by an SR554 transformer preamplifier. Both the voltage and current signals were then fed to a digital oscilloscope, which recorded VI characteristics of the samples. The transformer decoupled the current- and voltage- loops of the set-up, preventing the creation of additional current path through the common oscilloscope ground, in parallel to the current-path through the sample. Without this decoupling, it would not be possible to perform the measurements. Maximum achievable current with our set-up was typically 700 A. To check the reliability of this experimental set-up, a copper wire was measured with this method and with a standard DC method. These two types of measurements gave the same results to within 1.7%.

The sample was mounted on a sample holder containing a thermometer and placed in a continuous flow helium cryostat. The long-term temperature stability in the cryostat was better than 0.1K. A superconducting magnet produced magnetic field of up to 12 T. The field homogeneity in the sample area was 0.01%.

The temperature of the sample holder was monitored continuously during the measurements. It was noticed that, after the peak current of the pulse grossly exceeded the value of $I_c$, the temperature of the sample holder abruptly increased. For example, measurements were performed at 24K with peak current of 700A exceeding the $I_c$ by 200A. The temperature of the sample holder did not change for 5 seconds after the pulse, to suddenly jump by 0.1 K. This increase of temperature of 0.1K was dissipated gradually into the environment (i.e. cooling He gas) within the next few seconds. To check if the produced heat affected the measurements of $I_c$, the value of the peak current of the pulse was slightly decreased in each subsequent measurement. The increase of the temperature was not obtained any more as the peak current approached $I_c$ from above. Nevertheless, $I_c$ remained the same as with the peak current of 700A. As a further precaution, the value of the peak pulse current was always chosen to be within only a few tens of amperes above $I_c$.

Apparently, most of the heat was produced for $I > I_c$, when almost all of the current flew through the iron sheath, heating it up. However, the measurement was performed just after the current reached $I_c$ and this heat was very low. Further, the heat produced on the current contacts to the sample did not affect the measurements. This was because the voltage contacts were placed at a fair distance from the current contacts. By the time this heat reached the sample volume between the voltage contacts, the measurement has already been performed. Because of all this, the experimental results were not affected by the sample heating.

There was also a concern that the rate of change of the current could affect the experimental results. To check this, a programmable power source was used, which enabled synthesizing the current pulse of arbitrary shape and length, with maximum peak current of about 250 A. Here, the current was ramped to its peak value with a constant rate. The measured value of $I_c$ did not change as the ramp time needed to attain the 250 A changed from 0.3 millisecond to 50 milliseconds. This confirmed that the measurements with the capacitor as a current source did not depend on the rate of change of the current.

A typical VI characteristic obtained by the pulse method is shown in Figure 1. The initial step in the voltage occurred because of induction of voltages in the voltage loop on the sample, as the sample produced a time-dependent self-field upon the current pulse. There was also a contribution from the magnetization of the iron sheath by the self-field. Because these parasitic voltages were large, measurements of the voltages produced by the superconductor were severely limited in resolution. However, this did not affect the measurement of $I_c$, because the voltage for $MgB_2$ increased abruptly at $I = I_c$ (Fig.1).

**4. Field dependence of $J_c$**

The field dependence of $J_c$ for $MgB_2$ wires is usually measured by magnetic method. This is because the current contacts would heat too much in the transport measurements, due to the large value of $I_c$ of $MgB_2$ wires. Figure 2 shows the field dependence of $J_c$ normalized to its value in zero field for an $MgB_2$ wire, obtained from magnetic measurements. The iron sheath was removed, to enable the use of the critical state model[19] in calculating $J_c$. Otherwise, the sheath would screen external field from the superconductor, as well as the signal of the superconductor from the pick-up coils of the magnetometer[20]. As seen in Fig.2, the $J_c$ of the core extracted from an $MgB_2$/Fe wire decreases smoothly with the field.

The field dependence of $J_c$ for iron sheathed $MgB_2$ wire obtained from transport measurements is shown in Figure 3 for several temperatures. The applied field was perpendicular to the wire. The resistance of the current contacts in these measurements was minimised, because the current leads were soldered directly to the iron sheath, which was in good electrical contact with the superconducting core. Even so, the measurements had to be performed by the pulsed-current method. There is an apparent difference between the measurements in Figs 2 and 3. With the iron sheath, $J_c$ initially decreases with the field (Fig.3). This is followed by a plateau for intermediate fields and $T > 30$ K [16]. Finally, $J_c$ again decreases with field for high fields. However, the plateau in the field dependence of $J_c$ changes with the temperature. This plateau gradually develops into a peak as the temperature decreases below 30K (Fig.3). There, $J_c$ increases with the field, followed by a decrease in higher fields[17].

The field range where the plateau and peak occur gradually widens as the temperature decreases (Fig.3). To obtain a better insight into the temperature dependence of this field range, the field $H_p$ is defined as the field of the peak for $T < 30$ K, or as the field of the upper limit of the plateau for $T > 30$ K (Fig.3). Figure 4 shows the temperature dependence of $H_p$. $H_p$ increases from 0.9T at 27K to 3.4T at 10K. The increase of $H_p$ accelerates as the temperature decreases.

Figure 5 shows the field dependence of $J_c$ with the applied field parallel to the wire axis, therefore parallel to the overall current direction through the sample. This field dependence of $J_c$ is different than for the perpendicular field. Instead of the initial decrease of $J_c$, there is an increase of $J_c$, followed by a decrease for higher fields at T>30K. Below 30K, there is a plateau at low fields, followed by a faster decrease at high fields. Field $H_0$ is defined as the field of the transition from the plateau (or peak) to the field range of faster decrease of $J_c$ (Fig. 5). The temperature dependence of $H_0$ is shown in Figure 6. Similar to $H_p$, $H_0$ increases progressively stronger as the temperature decreases.

Figure 7 compares the field dependence of $J_c$ for parallel and perpendicular field configuration at 27K. For the parallel field, $J_c$ initially changes very little with the field. Thus, the value of $J_c$ is higher than for the perpendicular field, which initially decreases with the field. However, the peak occurs at higher fields for the perpendicular configuration. Because of this, $J_c$ becomes higher for the perpendicular field. This difference is strongly temperature dependent. At high temperatures, the difference in $J_c$ for the two field orientations is small at low fields, but large at high fields [16]. At temperatures lower than 30K, it is the opposite (Fig. 7).

Figure 8 shows the angle dependence of $I_c$ for an $MgB_2$/Fe wire at 33.7 K and 0.4 T. This corresponds to the high field regime, with large difference between $J_c$ for parallel and perpendicular field. $\theta$ is the angle between the field and the long axis of the wire. Therefore, $\theta = 90°$ corresponds to the perpendicular field. $I_c$ decreases by 75% of its maximum value within 30° from the perpendicular direction. For higher values of $\theta$, $I_c$ remains almost constant. This angle dependence of $I_c$ is quite different from the cosine-law, which would be expected if a projection of the field onto the perpendicular to the wire defined $I_c$.

It should be noted that the configuration with the field parallel to the wire is nominally the Lorentz force-free configuration. If the current flew through the wire in a straight path, the Lorentz force on the magnetic vortices due to the external field would be zero and field would not affect $J_c$[21, 22]. However, there is an apparent field dependence of $J_c$ for the parallel field (Fig.5). Moreover, the field dependence of $J_c$ for parallel and perpendicular field is the same for the high-field range (Fig.7). If the $J_c$ for the parallel field is shifted along the H-axis, its high- field part will overlap with the high-field part of the $J_c$ for the perpendicular field[16]. This implies that, in the high-field range, iron sheath does not affect $J_c$ any further and that the current meanders between the superconducting grains in the core in a random manner. The idea of meandering of the current is supported by low density of the superconducting core (65%) and by magnetic measurements showing different degrees of the coupling between the grains on different length-scales[23]. Because the local Lorentz force is proportional to the sine of the angle between the local current and field, the Lorentz force-free configuration is not achieved on the microscopic scale for the parallel field. Instead, Lorentz force varies locally, depending on the *local* direction of the current flow. As the variation of the current direction across the wire is random, the average Lorentz force is the same regardless the field direction. Therefore, the field dependence of $J_c$ would be the same for all $\theta$, if there were no iron sheath around $MgB_2$ core.

Comparing the field dependence of $J_c$ for bare $MgB_2$ superconductor (Fig. 2) with the iron sheathed $MgB_2$ (Figs 3 and 5), there is an apparent improvement for the iron sheathed wires. For the perpendicular field (Fig. 3), the decrease of $J_c$ in low fields is counteracted by the appearance of the peak at intermediate fields. For the parallel field, a plateau appears in low fields, improving the field dependence of $J_c$ over the bare superconducting core.

Strictly speaking, the magnetic and transport measurements are not the same. In magnetic measurements, the field was parallel to the wire axis and the current was flowing around the perimeter of the wire. In transport measurements, the current flew along the wire and the field was either parallel or perpendicular to it. Further, there was a self-field produced by the current in the transport measurements, which was oriented along the φ-axis of the natural cylindrical coordinate system of the wire, similar to the current in magnetic measurements. The voltage criterion for defining the $I_c$ is also different for transport and magnetic measurements.

All this can give a difference in the value of $J_c$ when comparing the transport with magnetic measurements. However, the field dependence of $J_c$ is still expected to be very similar for the two types of the measurements. Namely, the voltage increases very steeply at $I = I_c$ for $MgB_2$ over most of the field range and the voltage criterion does not affect the value of $I_c$ substantially. Further, the current meanders randomly inside the superconducting core and the average direction of the current and direction of the field should not affect the measurements substantially. Finally, the value of the self-field is much smaller than the value of $H_p$, and its influence on $I_c$ should be small, too. In the light of this, the magnetic measurements are expected to give very similar field dependence of $J_c$ to the one obtained from transport measurements.

This argument is in agreement with the magnetic measurements performed on the $MgB_2$/Fe wire with and without the iron sheath[24]. After subtracting the contribution of the magnetic hysteresis loop of the iron, the obtained field dependence of $J_c$ was better for $MgB_2$ with the iron sheath on it, than for the bare $MgB_2$ core. This confirms our findings that $J_c$ for the iron-sheathed $MgB_2$ is better than for the bare $MgB_2$ core.

## 5. Magnetic shielding by iron sheath

An apparent mechanism of improving the field dependence of $J_c$ by the iron sheath would be the magnetic shielding. Because the magnetic susceptibility of iron is much higher than of air, applied magnetic field is channelled through the iron sheath. In this way, little field reaches the superconductor, up to the field at which the iron approaches magnetic saturation. The fields beyond this are passed through the sheath with no further shielding.

To assess the influence of the shielding effect on the observed field dependence of $J_c$ (Figs. 3, 5, 7, 8), magnetic field was measured inside and outside a hollow cylinder of iron. The cylinder was prepared by drilling out the $MgB_2$ core from the sample of $MgB_2$/Fe wire. The cylinder was placed in the middle of a solenoid, driven by a 12kW AC power source. Both solenoid and cylinder were submerged in liquid nitrogen. The solenoid produced an AC magnetic field with maximum amplitude of 0.6 T and frequency 20-60 Hz. A tiny pick-up coil was placed inside the hollow cylinder, to measure the field inside it ($H_{in}$). Removing the cylinder, the voltage induced in the pick-up coil was proportional to the known field of the solenoid, $H_{out}$, which was used as a calibration.

The field was applied in either perpendicular or parallel direction to the cylindrical z-axis, in the same way as in the measurements of $J_c$. A different pick-up coil was used for each of the field directions, so that the pick-up coil was always parallel to the solenoid. The signal from the pick-up coil was captured by a digital oscilloscope, enabling instantaneous measurements of the field as it varied in time. By comparing the measurements at frequencies between 20 and 60 Hz, it was obtained that the dynamic effects did not play a role below 30Hz. The dimensions of the coils corresponded to the distance between the voltage contacts in the measurements of $J_c$. In this way, the obtained signal represented an average shielding effect over the whole volume of the sample that was contributing to the voltage in the measurements of $J_c$. Therefore, the effects of the finite wire length were taken into account.

Figure 9 shows the field measured inside the iron cylinder as a function of external field applied perpendicular to the cylinder z-axis. These measurements represent the magnetic shielding by the iron sheath for the measurements of $J_c$ in perpendicular field, shown in Fig. 3. For $H_{out} < 0.2T$, $H_{in}$ is negligibly small, reaching 0.008T at $H_{out} = 0.2$ T (open symbols in Fig 9). For $0.2\ T < H_{out} < 0.4\ T$, the field starts penetrating the cylinder and $H_{in}$ gradually increases. For $H_{out} > 0.4T$ the cylinder no longer shields any additional external field and $H_{in}$ vs. $H_{out}$ is parallel to the points measured without the iron cylinder (solid symbols in Fig. 9). The solid line in Fig. 9 shows the theoretical expression for the magnetic shielding with perpendicular field, for an infinitely long cylinder:[25]

$$H_{in} = \frac{H_{out}}{1 + \frac{(\mu-1)^2}{4\mu}\left[1 - \left(\frac{r_i}{r_o}\right)^2\right]}, \quad (1)$$

where $\mu$ is the magnetic permeability of the shielding material and $r_i$ and $r_o$ are the inner and outer radius of the cylinder, respectively. The difference between the theoretical curve and the results for low fields in Fig.9 probably occurs because of the finite size of the measured cylinder. For higher fields, the experimental points approach the theoretical curve.

The shielding effect of the iron cylinder for the field along the cylinder z-axis is shown in Fig. 10. This figure shows the magnetic shielding by iron sheath for the measurements of $J_c$ with parallel field shown in Fig. 5. The external field is completely shielded out, up to about $H_{out}$ = 0.02 T. The entire additional field above 0.02T is passed through the cylinder unhindered, and the measured points are parallel to the line $H_{in}$ = $H_{out}$. The small hysteresis in Fig. 10 occurs because of the magnetic hysteresis loop of the iron.

Comparing the Figures 3 and 5 with the corresponding measurements of magnetic shielding (Figs, 9 and 10, respectively), it is clear that magnetic shielding cannot account for the observed field dependence of $J_c$. Magnetic shielding would give a constant $J_c$ up to about 0.2 T and 0.02 T for perpendicular and parallel field, respectively. This would be followed by a decrease of $J_c$ as the entire field above these values is passed through the iron sheath. Further, magnetic properties of iron do not change almost at all in the temperature range 40 K > T > 20 K. Indeed the measured magnetic hysteresis loops of iron at 20 K and 40 K almost exactly overlap. Therefore, the shielding properties of the iron sheath remain the same for all the measured temperatures. This is in contrast to the measured field dependence of $J_c$. The width of the plateau in $J_c$ vs. H and the value of $H_p$ exhibit very strong temperature dependence (Figs. 3-7). Dashed lines in Figures 3 and 5 illustrate the expected field dependence of $J_c$ at 32 and 24 K, when the corresponding magnetic shielding shown in Figs. 9 and 10 is taken into account. Apparently, magnetic shielding is not the physical mechanism leading to the observed $J_c$ vs. H.

**6. Interaction of superconductor with magnetic environment**

Another way to explain the observed field dependence of $J_c$ is to take into account the interaction between the superconductor and magnetic sheath. A model predicting an increase of $J_c$ when a superconductor is placed in magnetic surrounding was proposed by Genenko et al.[26, 27]. In their model, the critical currents were calculated for type II superconducting thin strips in partly filled vortex state, in a magnetic surrounding. The magnetic surrounding was assumed to be reversible, linear, and homogeneous. Unlike the model of Majoros et al.[15], this model incorporates the influence of magnetic surrounding on the current distribution in the central vortex-free part of the strip. Only the self-field of the transport current was considered in the model, with no magnetic shielding from external field. The current distribution was found to be very sensitive to the shape of the magnetic surrounding and its distance to the superconducting strip. They predicted an increase of the maximum loss-free current by a factor of 100 when a superconducting strip is placed in an open magnetic cavity and by a factor 7 in practically achievable situations. This increase of the apparent critical current occurred because of re-distribution of the current into the vortex-free region of the superconducting strip due to the magnetic surrounding, whilst the vortices were confined to the edges of the strip by the edge barriers. In this way, an overcritical state is formed, with maximum loss-less current density larger than the critical state $J_c$.

Strictly speaking, this picture does not apply to the round wire. However, there are bulk pinning centres in the real samples, which may play the role equivalent to the edge barriers in the thin strip. Therefore, the basic mechanisms of the model may be valid for the round wires, too. Unfortunately, the model does not consider the influence of the external magnetic field on $J_c$ and its predictions cannot be compared with the experimental results presented here. However, it points to a viable physical background that could explain these results.

Another type of interaction between the superconductor and magnetic environment is the pinning of magnetic vortices through the interaction with their mirror image in the magnetic medium[28]. The mirror image of a vortex is of the opposite polarity and the vortex will be attracted to it. In this way, the vortices will be effectively pinned at the boundary of the superconductor. The field dependence of this pinning is affected by the change of magnetic properties of the magnetic environment with the field, as well as by the increased interaction between the vortices in the superconductor, as the field increases. At this stage, there is no model available for the field dependence of $J_c$ taking into account this type of interaction.

In the concept of lowering the AC loss by magnetic decoupling of the filaments, only magnetic shielding is considered[15]. However, the shielding is apparently of secondary importance in the measurements presented here. Because AC loss depends on the distribution of the currents in the superconductor, it is likely that the interaction between the superconductor and iron sheath will play a major role in the AC loss of $MgB_2$/Fe superconductor, too.

### 7. Vortex pinning

Vortex pinning was expected to have a strong influence on the plateau and peak shown in Figs, 3 and 5. Even though the above model [26, 27] does not show it explicitly, it is intuitively expected that the improvement of the vortex pinning will change the distribution of the vortices, which will in turn change the re-distribution of the current into the vortex-free region by magnetic sheath.

To check this hypothesis, vortex pinning in $MgB_2$/Fe wires was improved by doping with nano-size SiC particles [6]. This procedure produces very strong improvement of the field dependence of $J_c$. The same measurements as for the pure $MgB_2$/Fe wires were performed on the wires doped with 10 wt. % of SiC. The field dependence of $J_c$ at the highest fields was strongly improved. However, the values of $H_p$ remained the same as before the doping. This is shown in Fig. 4, where the solid and open symbols represent $H_p$ for the doped and pure wires, respectively.

Even though $H_p$ seems to be a natural parameter for describing the effects of iron sheath on $J_c$, it may not be suitable for studying the effects of pinning. There is no model available describing the field dependence of $J_c$ shown in Figs. 3 and 5, and therefore $H_p$. Actual mechanisms for the occurrence of $H_p$ are still not clear and the presented results may help elucidate these mechanisms.

### 8. Conclusions

A substantial improvement of the field dependence of $J_c$ was obtained for $MgB_2$ wires by using a magnetic sheath around the superconducting core. This improvement is much stronger and it extends to much higher fields, than achievable by a simple magnetic shielding of the superconducting core by the iron sheath. The magnetic shielding is effective up to the fields of 0.2 T and it is not temperature dependent in the measured temperature range, 40 K > T > 20 K. However, the improvement of $J_c$ occurs in the form of a plateau and peak in the field dependence of $J_c$. The plateau and peak are strongly temperature dependent. They extend to higher fields as the temperature decreases, way beyond the fields in which magnetic shielding is still effective.

Apparently, the observed phenomena cannot be explained by magnetic shielding. A more plausible explanation would be in the terms of interaction between the superconductor and iron sheath. Such interaction is predicted to increase the value of zero-field $J_c$, however there are no models available describing the effect of this interaction on the field dependence of $J_c$.

Figure 1: A typical voltage-current curve for $MgB_2$/Fe wire obtained by a pulsed current method [17].

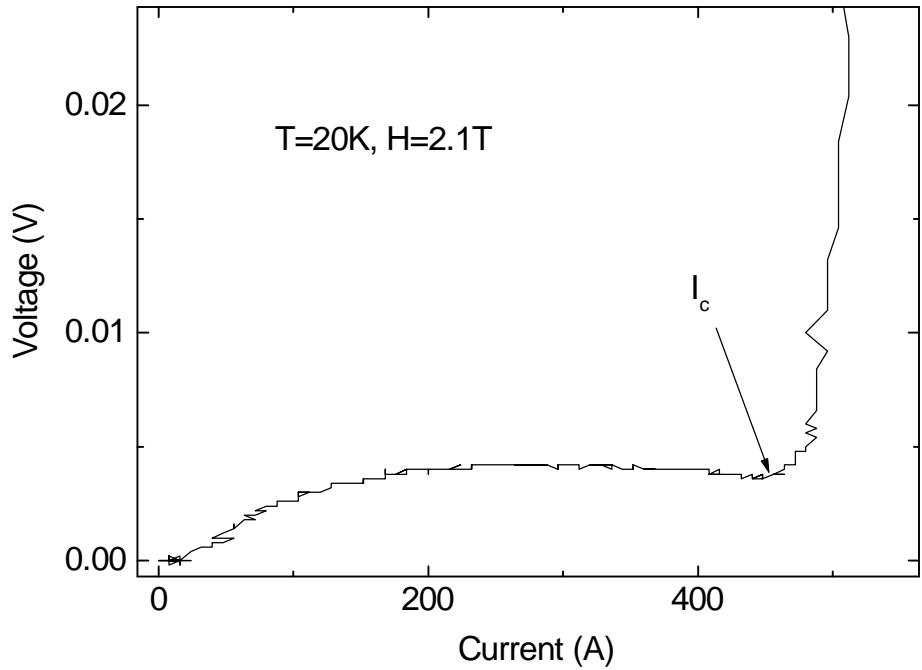

Figure 2: Field dependence of $J_c$ for $MgB_2$/Fe wire, obtained from measurements of magnetic hysteresis loop. Iron sheath was removed before the measurements.

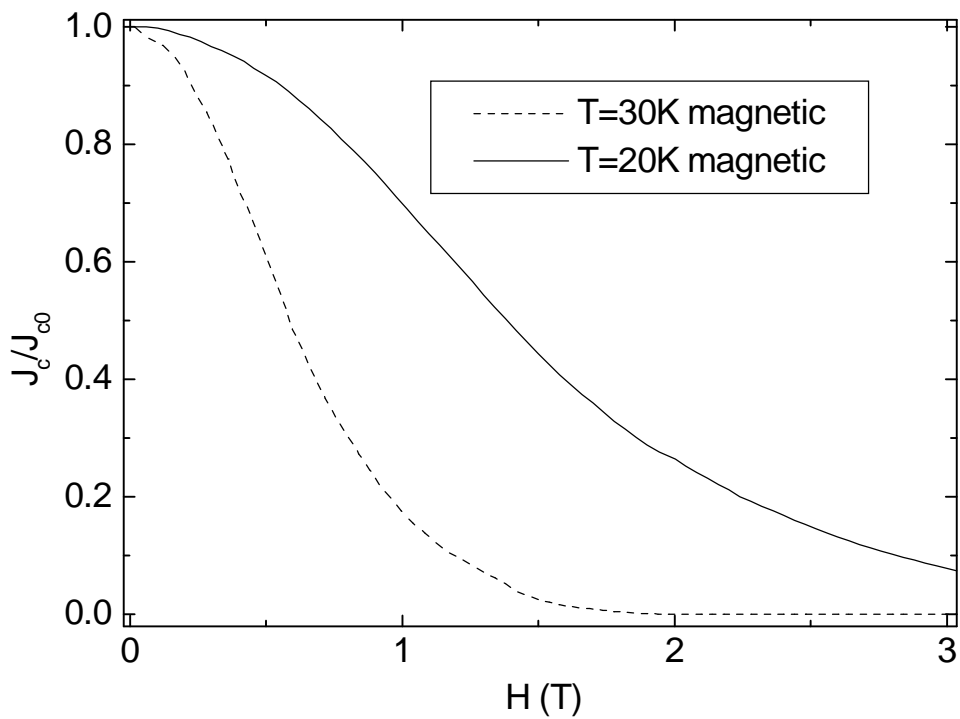

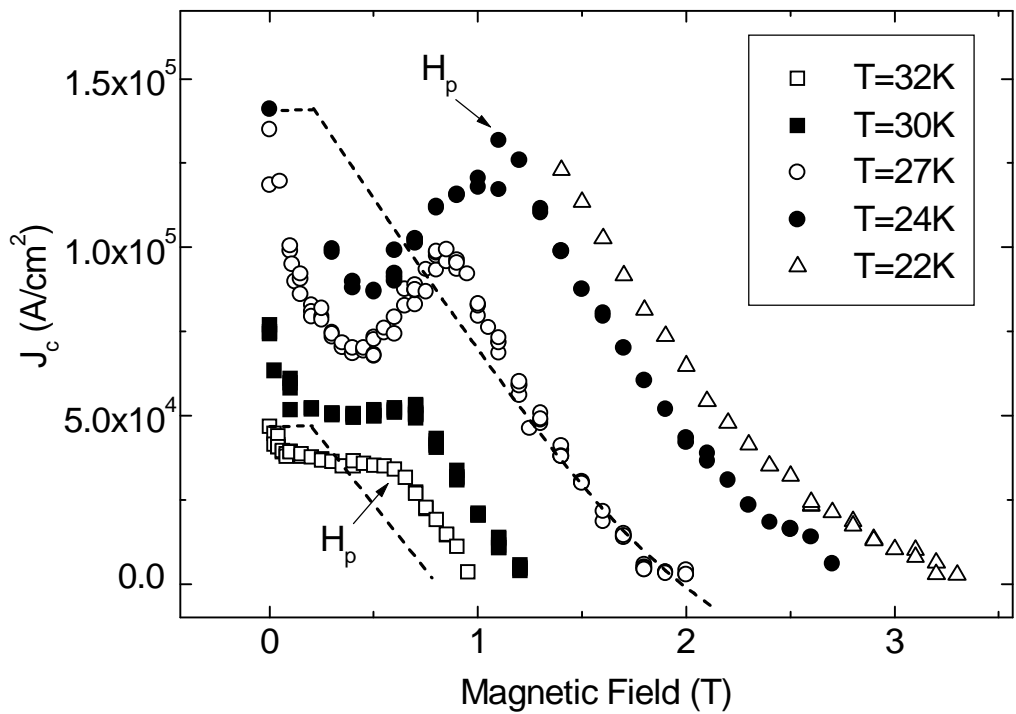

Figure 3: Field dependence of $J_c$ for $MgB_2$/Fe wire with perpendicular field. Dashed lines illustrate the anticipated field dependence of $J_c$ at 24 and 32 K, when the effect of magnetic shielding is taken into account [17].

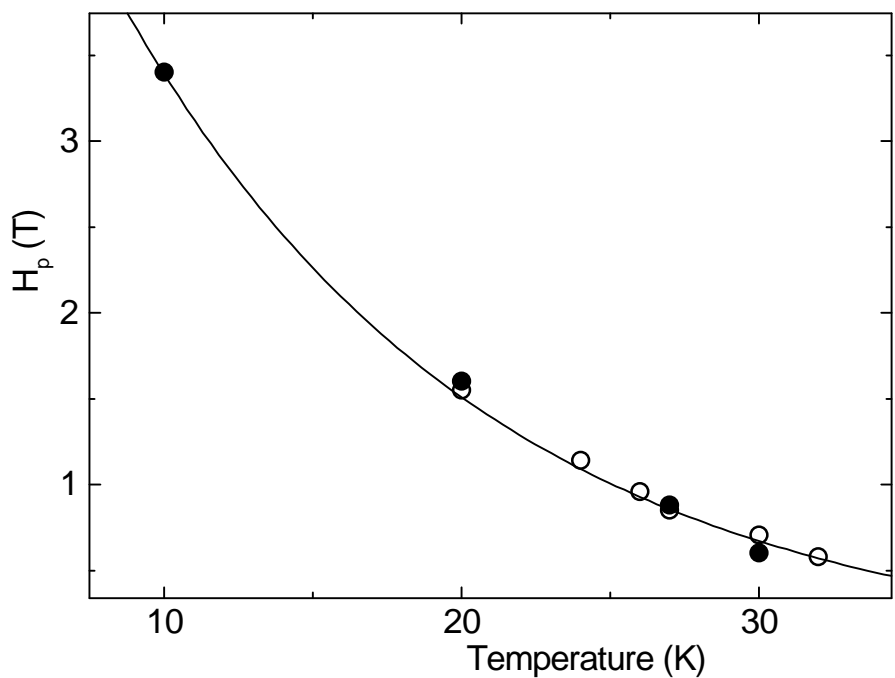

Figure 4: Temperature dependence of the field of the peak in the field dependence of $J_c$, for perpendicular field [17]. Solid and open symbols are for the nano-SiC doped and pure $MgB_2$, respectively.

Figure 5: Field dependence of $J_c$ for $MgB_2$/Fe wire with parallel field. Dashed lines illustrate the anticipated field dependence of $J_c$ at 24 and 32 K, when the effect of magnetic shielding is taken into account [17].

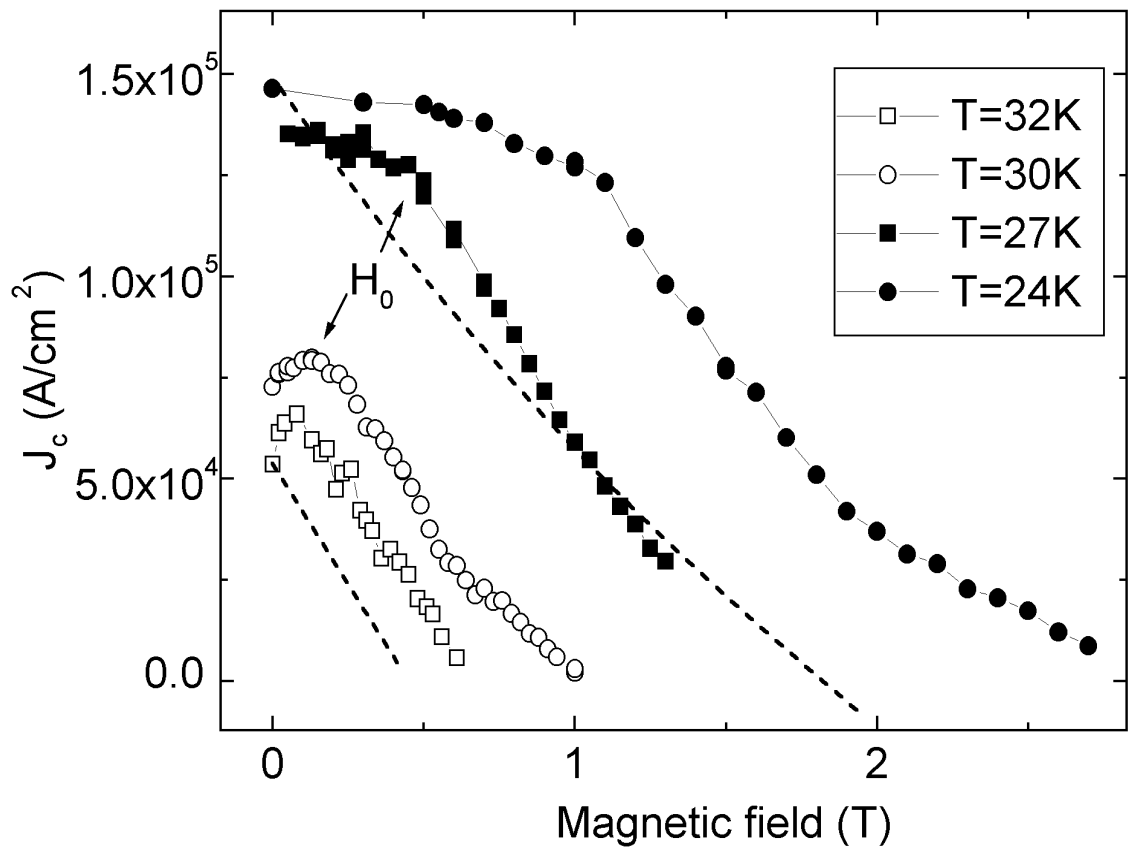

Figure 6: Temperature dependence of the field of transition from the plateau to decreasing regime in $J_c$ vs. H, for parallel field [17].

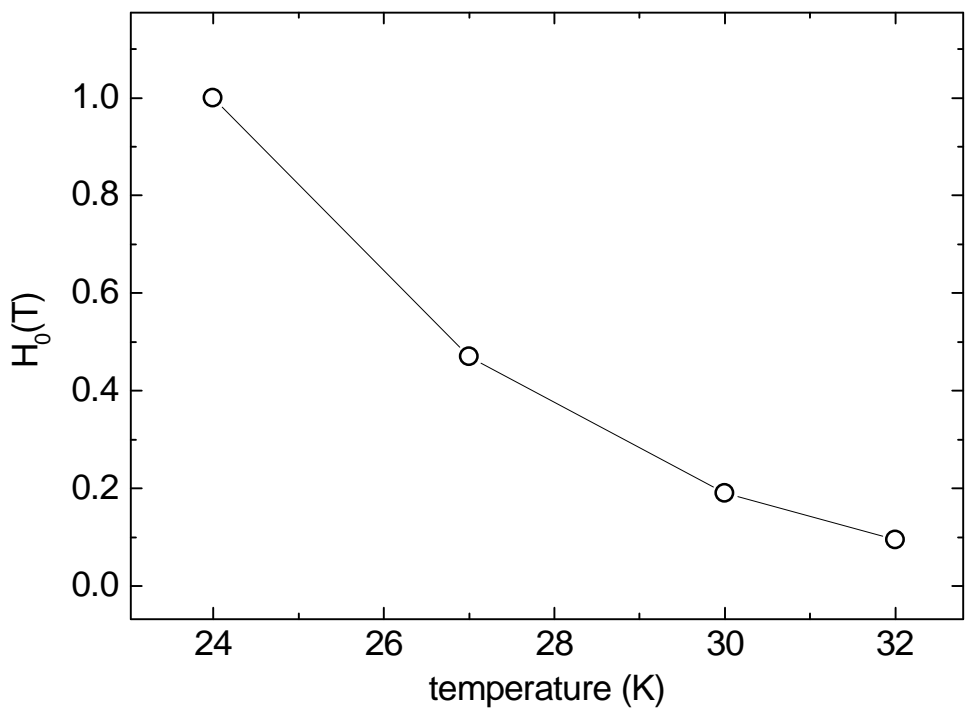

Figure 7: Comparison between the field dependence of $J_c$ for parallel and perpendicular field [17].

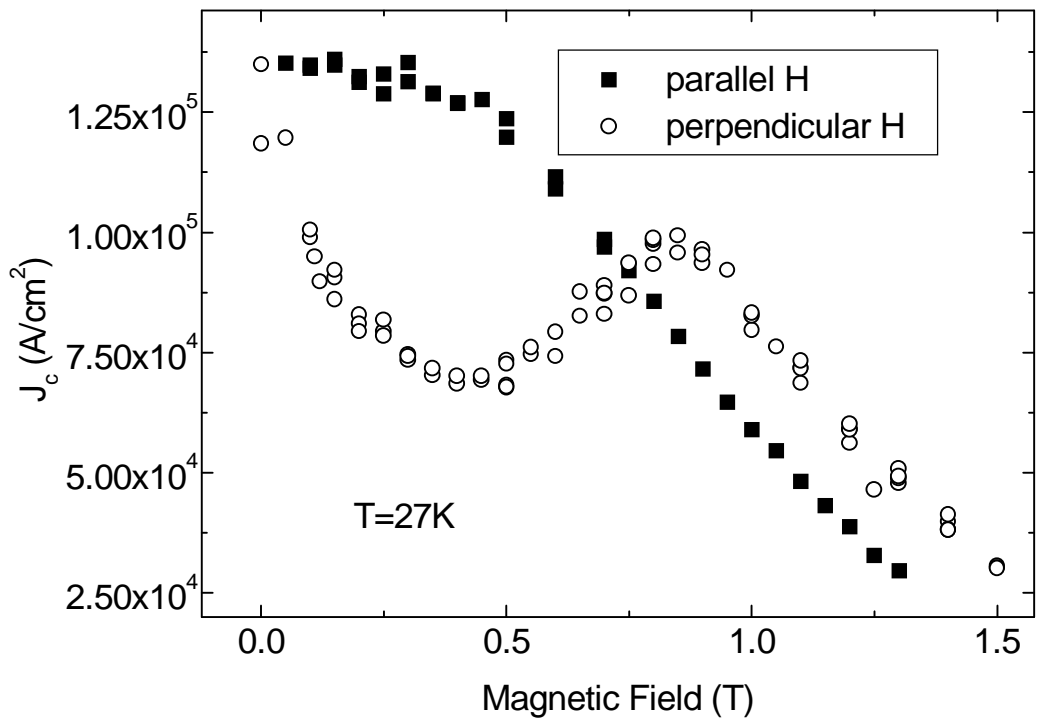

Figure 8: Angular dependence of critical current for $MgB_2$/Fe wire at temperature 33.7K and field 0.4 T [16].

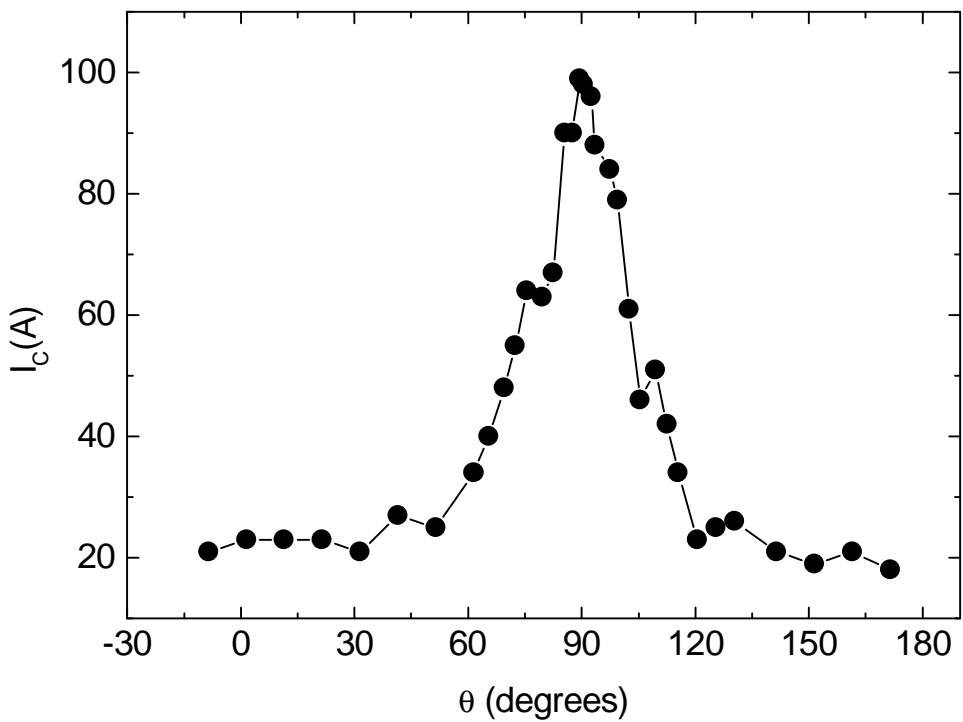

Figure 9: The magnetic field inside the iron sheath, $H_{in}$, against the external field, $H_{out}$, for perpendicular field (open symbols) [16]. With iron sheath removed, $H_{in} = H_{out}$ (solid symbols). Solid line shows the theoretically obtained magnetic shielding.

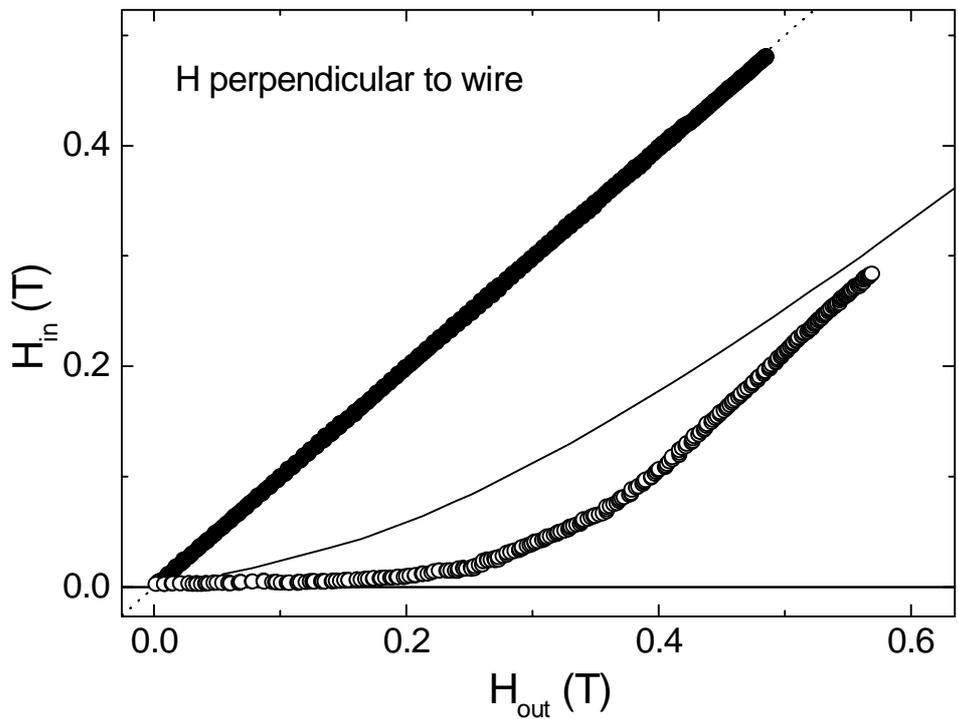

Figure 10: The magnetic field inside the iron sheath, $H_{in}$, against the external field, $H_{out}$, for parallel field (open symbols) [16].

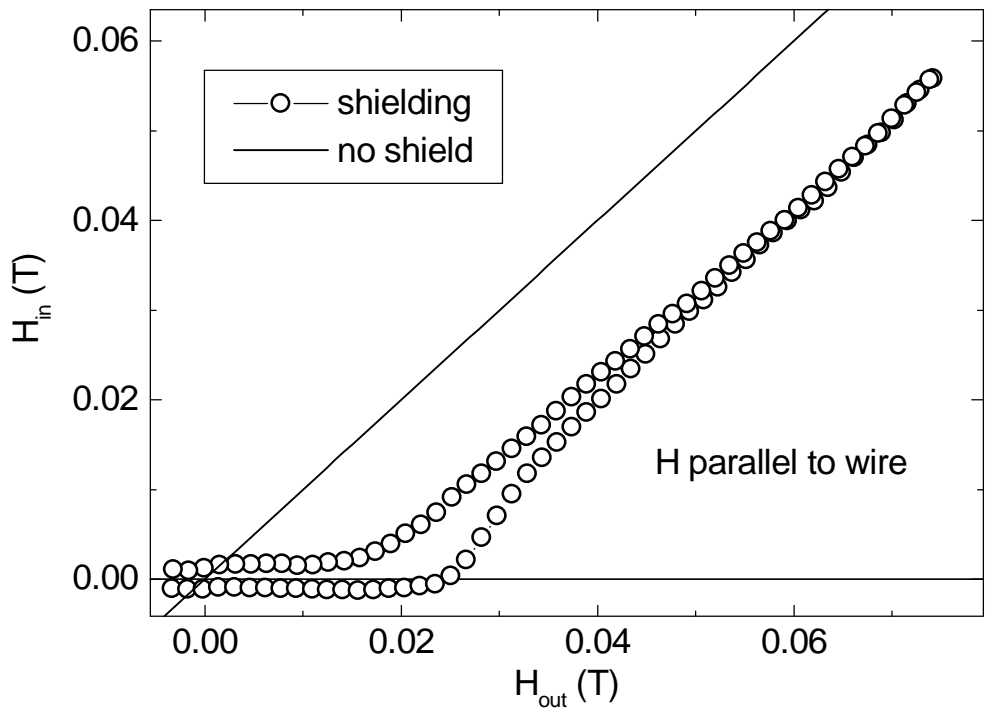